\documentstyle[12pt,doublespace]{article}
\def\Jparallel{\rm J_{\parallel}}
\def\tperp{\rm t_{\perp}}

\begin{document}
\title{\protect\large \bf Phase separation and pairing in coupled chains
and planes}

\author{Jos\'{e} A. Riera \\
\vspace{-3 mm}
\protect\small \em Center for Computationally Intensive Physics \\
\vspace{-3 mm}
\protect\small \em Physics Division, Oak Ridge National Laboratory, \\
\vspace{-3 mm}
\protect\small \em Oak Ridge, Tennessee 37831-6373  \\
\vspace{-3 mm}
\protect\small \em and  \\
\vspace{-3 mm}
\protect\small \em Department of Physics $\&$ Astronomy  \\
\vspace{-3 mm}
\protect\small \em Vanderbilt University     \\
\vspace{-3 mm}
\protect\small \em Nashville, Tennessee 37235}
\date{}
\maketitle

\begin{abstract}
A generalization of the $t-J$ model in a system of two coupled chains
or planes is studied by numerical diagonalization of small clusters.
In particular, the effect of density fluctuations between these one- or
two-dimensional coupled layers
on intralayer phase separation and pairing is analyzed.
The most robust signals of superconductivity are found at
quarter filling for couplings just before the fully interlayer
phase
separated regime.
The possibility of
an enhancement of the intralayer superconducting pairing correlations
by the interlayer couplings is investigated.

\end{abstract}

\vskip 0.5cm


\newpage

\hspace{2em}High-T$_c$ superconductors$^1$
have a layered structure consisting of Cu-O layers and some other
intermediate layers. While it is widely believed that the pairing
takes place within Cu-O layers, it is also well known that the
critical temperature depends also on the number of these layers
per unit cell and on the distance between them.$^2$
Consequently, since the early days of high-T$_c$ superconductivity,
there have been attempts to include these two features in a
consistent theoretical framework.$^3$

One of the simplest models proposed to describe Cu-O planes in
high-T$_c$ superconductors is the two-dimensional $ t - J $ model.$^4$
This model gives results in reasonable
agreement with experiments for many of the magnetic and
electronic properties of these materials.$^5$
Numerical studies have also given indications of pairing  in the
$d_{x^2-y^2}$ mode in some regions of the parameter space,$^{6,7}$
although in general, and especially for the physically relevant
region of parameters,
these indications are not the strong signals
we would expect that correspond to high-T$_c$ superconductivity.
This somewhat negative
behavior has led some authors to think that there is an important
feature which is missing in this model. Since there is a relation
between the number of planes and the value of the critical temperature,
it is natural to ask if the $t-J$ model could be generalized to include
the coupling between Cu-O layers.

Another important property that characterizes the phase diagram
of the $t - J$ model is the property of phase separation
(PS).$^8$
For sufficiently large values of $J/t$ the system undergoes a separation
of two phases: a hole-rich phase and an electron-rich phase. This
phase separation is driven by the same force that also gives rise to
pairing, which is a necessary condition for superconductivity.
It is possible that just before phase separation, density fluctuations
could enhance pairing leading to superconductivity. This scenario
has been confirmed in at least two situations: the $t-J-V$ model,
where $V$ is a repulsive interaction between holes in nearest-neighbor
sites,$^{9,10}$
and the two-dimensional (2d) $t-J$ model close to quarter
filling.$^6$
Another case we could add to this list is the one-dimensional (1d) $t-J$
model for $J/t$ larger than 2 (the ``supersymmetric" point).$^{11}$
In a more general study, Emery and Kivelson$^{12}$ suggested
that the property of phase separation could account for
some of the ``anomalous" properties that characterize the normal
state of high-T$_c$ superconductors.

In this sense, the first motivation for this work is to study
how interlayer coupling affects pairing correlations and phase
separation in the $t-J$ model in a system of two coupled layers.
In general, we refer to chains or planes
as one- or two-dimensional ``layers", respectively.

However, if the number of particles and total magnetization is kept fixed
for the whole system of coupled layers, a  new important
feature appears.
It is natural to assume that by symmetry there is an equal number of
holes on each chain or plane. However, this is only true {\em on
average}. As it is shown below, in many cases it is energetically
possible that there are {\em fluctuations} in the density of holes
between the two parts of the system. In this case, snapshots of the
system would show that there are {\em different} numbers of holes on
each of the layers. We call this situation ``interlayer
phase-separated regime".
This regime first appears in the absence of transversal
couplings, and in this case it is quite easy to understand its origin.

The main purpose of this work then is to study how the interlayer
PS evolves when transversal couplings are switched on,
and to study the interplay between this type of phase separation with
the intralayer PS and with intralayer pairing of holes.
In particular, we are interested in investigating previous ideas
about the relation of phase separation and superconductivity$^{12}$
taking into account this new source of density fluctuations.

For that purpose, we have used Lanczos techniques to calculate
the ground state in finite clusters.
Most of the results reported in this work refer to coupled
chains because in this case we can study correlations at larger
distances than in the case of coupled planes, and then we expect
to reduce finite size effects. Moreover, as can be seen below,
most of the properties studied show a very similar behavior
for coupled one- and two-dimensional layers.

The $t-J$ model for the two-layer system is
defined by the Hamiltonian:
\begin{eqnarray}
H = - t_{\parallel} \sum_{\scriptstyle <i j>, \alpha, \sigma}
(\tilde{c}^{\dagger}_{\alpha, i,\sigma}\tilde{c}_{\alpha, j,\sigma} +
\tilde{c}^{\dagger}_{\alpha, j,\sigma}\tilde{c}_{\alpha, i,\sigma})
+ J_{\parallel} \sum_{ <i j>}
({\bf S}_{\alpha, i}\cdot {\bf S}_{\alpha, j}
- \frac{1}{4} n_{\alpha, i} n_{\alpha, j})    \\
\nonumber
- t_{\perp} \sum_{\scriptstyle i, \sigma}
(\tilde{c}^{\dagger}_{1, i,\sigma}\tilde{c}_{2, i,\sigma} +
\tilde{c}^{\dagger}_{2, i,\sigma}\tilde{c}_{1, i,\sigma})
+ J_{\perp} \sum_{ i}
({\bf S}_{1, i}\cdot {\bf S}_{2, i} - \frac{1}{4} n_{1, i} n_{2, i}),
\end{eqnarray}

\noindent
where the notation is standard. The label $\alpha =1,2$ indicates the
two layers of the system. The couplings ($t_{\parallel}, J_{\parallel}$)
refer to the intralayer interactions, while ($t_{\perp}, J_{\perp}$)
are the interlayer coupling constants.
In the case of two coupled chains, most of our results are for
the 8 sites chain with periodic boundary conditions.
In the case of coupled planes, we have considered for each plane the
tilted cluster denoted as $\sqrt{8} \times \sqrt{8}$.
The total number of sites in both lattices is
$N_s = 16$.
For each hole density, $x = N_h / N_s$ ($N_h$ : number of holes),
we have considered $t_{\perp}$ = 0.0, 0.2, 0.4, 0.6, 0.8, and 1.0.
As usual, we adopt $t_{\parallel} = 1$.
For each $\tperp$, we have taken $J_{\perp} = J_{\parallel} \times
t_{\perp}^2/t_{\parallel}^2$, and $\Jparallel$ remains as a
free parameter.

This relation between $J_{\perp}, J_{\parallel}, t_{\perp}$, and
$t_{\parallel}$ is the one that would appear from the reduction of
the one- or three-band Hubbard model to an effective $t - J$
model, in the limit of strong Coulomb repulsion in the copper
sites.$^{13}$
However, we are going to consider
values of $J_{\perp}, J_{\parallel} >> t_{\parallel}, \tperp$, i.e.
beyond the range of validity of such construction.
The main reason to adopt that relation is to recover the isotropic
situation when $t_{\perp} = 1$.$^{14}$
Presumably
a value of $J_{\perp}$ constant$^{15}$ and eventually equal
to zero$^{16}$ should be considered$^{17}$.

This model has been numerically examined before for the case $J_{\perp}
= 0$ $^{16}$ although for
rather smaller lattices. In that study, it was found that intralayer
superconducting correlations are enhanced by $t_{\perp}$. In the
present study we obtain different results, as discussed below.

The fluctuations of hole density between layers or interlayer phase
separation are easy to understand in the case of the uncoupled
$t_{\perp} = 0$ limit.
In this case we can study larger clusters in order to avoid possible
finite size effects. Let us consider the case of two $4 \times 4$
planes with a total number of 16 holes (quarter filling). For
$J_{\parallel} = 0$, the ground state of the total system consists
of the product of the ground state of each plane with 8 holes each,
i.e. the total energy is $E(J_{\parallel} = 0) = E_1^{(8)} +
E_2^{(8)} = 2 \times (-14.3475)$. In this case there are no density
fluctuations and we call this regime an interlayer homogeneous regime.
Now, for $E(J_{\parallel} = 2)$, the ground state consists
of two parts: one which is a product of the ground state in one plane
with 6 holes and the ground state on the other plane with 10 holes,
and the other in which the number of holes in the two planes are
interchanged. In this case, $E(J_{\parallel} = 2) = E_1^{(6)} +
E_2^{(10)} = -26.0627 + (-17.4066) < 2 \times E_1^{(8)} = 2 \times
(-21.6386)$.
In the same way, at $J_{\parallel} \geq 3$, the system jumps back and
forth between the state in which one layer is empty and the other is at
half-filling and the state where these occupancies are interchanged.
It is worth noting that for $J \simeq 3.5$, at quarter filling, a
previous study$^6$ found the peak of the pairing susceptibility
with long-range pairing correlations in the $4 \times 4$ cluster.
We see now that for two coupled $4 \times 4$ planes, this density is
not stable at this value of $J$. Similar behavior is also found
for two chains of 16 sites each in the $t_{\perp} = 0$ case, and
for various fillings.$^{18}$

To study these interlayer density fluctuations, the natural order
parameter is then $f_d(J_{\parallel}, t_{\perp},x) = <N_1^2> - <N_1>^2$,
where $N_1$ is the hole number operator for one of the layers,
and $<N_1> = N_h / 2$ by symmetry.
This order parameter varies between zero, which corresponds to the
interlayer homogeneous state, and a number which depends on the
number of sites and on the number of holes, which corresponds to the
fully phase-separated interlayer regime. The same information is given
by the sum of the hole-hole correlations over the sites of one
layer.

The second property we are interested in measuring is the intralayer
PS. There are several ways to measure this property.
For the case of coupled chains, we have computed a very
direct measure of the clustering of holes defined as$^{19}$
$X = S(2 \pi / L)$, where $S$ is the Fourier transform of the
hole-hole correlations along one chain and $L$ is the chain length.
This quantity varies between zero (even number of holes) for
the homogeneous phase, and a positive number which can be calculated
analytically when the holes are clustered together. This
definition of $X$ can be extended to 2d.

To study superconductivity, we compute the pairing correlations:
\begin{eqnarray}
C(j) = \frac{1}{L} \sum_{\scriptstyle i} < \Delta_{i+j}^{\dagger}
\Delta_{i} >
\end{eqnarray}

\noindent
where the pairing operator $\Delta_i =  \sum_{\scriptstyle \mu}
g_{\mu} c_{1,i+\mu,\downarrow} c_{1,i,\uparrow}$, $i+\mu$ are the
nearest-neighbors sites of site $i$.
For chains $g_{\mu} = 1$ for $\mu=1,2$. For planes, $g_{\mu}$ defines
the extended s or the $d_{x^2-y^2}$ symmetries in the usual way$^{6}$.
The pairing susceptibility is defined as
$\chi_p = (1/L) \sum_{\scriptstyle j} C(j)$,
where the sum extends over the sites of {\em one} layer.

We start by discussing the results for coupled chains of 8 sites
each with 4 holes. Figure 1a shows the interlayer density fluctuations
as a function of $J_{\parallel}$. For $t_{\perp} \leq 0.6$ there is
a sharp crossover from the homogeneous to the phase-separated regime
at $J_{\parallel} \simeq 4.25$. In the PS regime the number of holes
in each plane is 0 or 4. For $\tperp \geq 0.8$ this crossover
disappears and the system is homogeneous for all $\Jparallel$.
Figure 1b shows the intralayer order parameter X as a function of
$\Jparallel$ for the same values of $\tperp$. In this case, X approaches
its maximum value for 4 holes, equal to its maximum value for 2
holes, $X_{lim} = 1.707$. For small values of $\tperp$ there is a
sharp increase for the value of $\Jparallel$ at which the interlayer
PS occurs. The pairing susceptibility as a function of $\Jparallel$
is shown in Fig. 1c. For $\tperp \leq 0.4$ the peak of $\chi_p$
is located just before the interlayer PS border, and the decrease
of $\chi_p$ is very sharp as it enters in this region.
By contrast, for $\tperp \geq 0.6$, the peak of $\chi_p$ starts
to shift to lower values of $\Jparallel$ and its intensity decreases.
Figure 1d shows the pairing correlations as a function of the
distance for the values of $\Jparallel$ at which $\chi_p$ has its
peak for each value of $\tperp$. At these values of $\Jparallel$ the
pairing correlations at the largest distance, C(L/2), has its
largest value for all $\Jparallel$ at a given $\tperp$. It is seen
that the decrease in the peak of $\chi_p$ as $\tperp$ is
increased corresponds essentially to a reduction of C(r) at
short distances. In fact, for $\tperp = 1.0$ we actually see
a tiny enhancement of C(L/2) with respect to $\tperp = 0$, but in any
case C(L/2) is too small to indicate the presence of superconductivity.

The same properties were computed
for the same system but with
$J_{\perp} = 0.0$ for all $\tperp$.
In general their behavior is quite similar to those shown in Fig. 1a.
The main difference found was that, for the interlayer density
fluctuations, as $\tperp$ increases, the crossover
from interlayer homogenous to PS regimes takes place for smaller
values of $\Jparallel$ as $\tperp$ increases, and that even for the
largest value of $\tperp$ studied, this crossover is still present
although somewhat rounded off.
At the filling considered (4 holes),
the pairing correlations decay also very rapidly as a function
of the distance.

Some of these patterns are also found for the case of quarter filling
for the same lattice. However, at this filling the processes
are more interesting. In the absence of coupling between chains,
$\tperp =0$, the system undergoes two successive interlayer
crossovers (Fig. 2a), first from the homogeneous state
to the state where
there are 2 or 6 holes on each chain indicating a partial
PS, and the second to a state in which there are 0 or 8 holes
on each chain, or fully interlayer PS. As $\tperp$ is increased,
the first crossover becomes smoother, while the second is
still sharp up to $\tperp \simeq 0.4$. Finally, for $\tperp \geq
0.6$, the second crossover is also washed out. For $\tperp \geq
0.8$, the system stays in the interlayer homogeneous state
for all $\Jparallel$.
For large $\tperp$, the behavior of the intralayer PS order
parameter (Fig. 2b) is very similar to that of Fig. 1b, where for
large $\Jparallel$, X approaches its maximum value $\simeq 1.707$.
However, for small $\tperp$, it can be seen that X increases
with $\Jparallel$ but has a {\em decrease} each time the system
has a crossover to an interlayer PS state. This behavior in
X can be understood by computing the maximum
values it can take for the system with (4,4) holes quoted
above, the system with (2,6) holes equal to 0.854, and for
the fully PS regime (0,8), where $X = 0$.
Figure 2c shows the intralayer pairing susceptibility $\chi_p$,
which, as in the case of 4 holes, presents its maximum
in the region of partial interlayer PS
close to the crossover to the fully phase-separated interlayer
regime for
$\tperp \leq 0.4$. Also, as in the case of 4 holes for large
$\tperp$, the peak is shifted to smaller values of $\Jparallel$
and its intensity decreases. The pairing correlations as a
function of distance, computed at the peak of $\chi_p$,
are displayed in Fig. 2d. First, it must be
emphasized that in this case the correlations
show a much more robust behavior at large distance than in the case
of 4 holes. This behavior is reminiscent of similar results
in one and two dimensions for the $t-J$ model,$^{6,10}$
indicating the presence of superconductivity.
Secondly, it is clearly seen that the pairing
correlations are {\em suppressed} by the interlayer
coupling. Another possibility of examining these results
consists in varying $\tperp$ keeping $\Jparallel$ fixed. For
example, the pairing correlations computed at
$\Jparallel = 2.25$ show the most robust
pairing at large distances at $\tperp \simeq 0.8$.
A similar enhancement of the pairing correlations for nonzero
interlayer coupling occurs at $\Jparallel = 5.0 $.
A detailed analysis of this behavior is given in Ref. 17.

Finally, in Fig. 3 we show some results for two coupled
$\sqrt{8} \times \sqrt{8}$ planes. We consider periodic boundary
conditions along both directions in the plane.
As in the case of two coupled chains, we found the strongest
signal of superconductivity at quarter filling. Fig. 3a
shows the interlayer density fluctuations as a function of
$\Jparallel$ for various values of $\tperp$. The two
crossovers found are equivalent to those found in the
coupled chains case at the same filling. However, even for
$\tperp = 1.0$ there is in this cluster a crossover to
the interlayer phase-separated regime. The intralayer
order parameter $X$ also shows the typical vanishing as
$\Jparallel$ increases beyond the region of fully PS.
The most important feature, which generalizes to coupled
2d systems what was already found for coupled 1d systems,
is the presence of the peak of the pairing susceptibility
in the region of partial interlayer PS just before
the crossover to the fully PS regime. At these
values of $\Jparallel$, the s-wave pairing correlations have
a true long-distance behavior, as can be seen for $\tperp =
0.0, \:0.6, and \:1.0$ in Fig. 3d. In this figure we have included
for comparison the pairing correlations for the 4 holes
case, computed also at the maximum of the pairing
susceptibility. The fact that s-pairing dominates over
d-pairing, contrary to what was found for the $4 \times 4$
lattice,$^6$ is presumably a finite size effect.

Summarizing,
the idea of enhancement of pairing close to phase separation,
as exposed in previous studies$^{10,11,12}$, is verified in a
new scenario. The new feature of the present study with respect
to previous work on the $t - J$ model is the presence of
a different source of density fluctuations: the interlayer
phase separation phenomena. For a fixed value of $\tperp$,
the strongest pairing occurs, as $\Jparallel$ is varied,
{\em just before} the crossover to the fully interlayer PS regime.
In general, the pairing is suppressed by interlayer coupling.
However, in some cases, if $\Jparallel$ is kept fixed,
pairing correlations are {\em enhanced} by increasing the
interlayer coupling, according to previous results for the
$t - J$$^{15}$ and Hubbard models.$^{20}$

\vspace{0.5 cm}
{\large \bf Acknowledgements}

The author wishes to acknowledge E. Dagotto for many discussions
on related problems, and to A. Dobry, M. Kovarik and J. Sofo for
useful comments.
This research has been supported in part by the U. S. Department
of Energy (DOE) Office of Scientific Computing under the
High Performance Computing and Communications Program (HPCC),
as a Grand Challenge titled the Quantum Structure of Matter,
and in part by DOE under contract No. DE-AC05-84OR21400 managed by
Martin Marietta Energy Systems, Inc., and under contract No.
DE-FG05-87ER40376 with Vanderbilt University. Most of the
calculations were done using the Cray YMP at the
Supercomputer Computations Research Institute in Tallahassee,
Florida and the Cray 2 at the National Supercomputer Center in
Urbana, Illinois.

\newpage

{\large \bf References}

\begin{enumerate}
\item J. G. Bednorz and K. M\"uller, {\em Z. Phys.} {\bf B 64}, 189 (1986).
\item H. Nobumasa, K. Shimizu, M. Nishina and T. Kawai, {\em Z. Phys. }
{\bf B 90}, 387 (1993), and references therein.
\item See for example, J. M. Wheatley, T. C. Hsu and P.W. Anderson,
{\em Phys. Rev.} {\bf B 37}, 5897 (1988).
\item P. W. Anderson, {\em Science} {\bf 235}, 1196 (1987).
\item E. Dagotto, A. Moreo, F. Ortolani, D. Poilblanc
and J. Riera, {\em Phys. Rev. } {\bf B 45}, 10741 (1992).
\item E. Dagotto and J. Riera, {\em Phys. Rev. Lett.} {\bf 70},
682 (1993).
\item D. Poilblanc, J. Riera, and E. Dagotto, preprint, (1993).
\item V. J. Emery, S. A. Kivelson, H-Q. Lin, {\em Phys. Rev. Lett.} {\bf
64}, 475 (1990).
\item S. A. Kivelson, V. J. Emery, H-Q. Lin, {\em Phys. Rev. } {\bf B 42},
6523 (1990).
\item E. Dagotto and J. Riera, {\em Phys. Rev. } {\bf B 46}, 12084 (1992);
T. Barnes, D. Jacobs, M. Kovarik and J. Macready, in preparation.
\item M. Ogata, M. U. Luchini, S. Sorella and F. F. Assaad, {\em Phys. Rev.
Lett.} {\bf 66}, 2388 (1991).
\item V. J. Emery and S. A. Kivelson, {\em Physica} {\bf C 209}, 597
(1993).
\item F. Zhang and T. M. Rice, {\em Phys. Rev.} {\bf B 37}, 3759 (1988).
\item The physical relevance of this limit
is discussed in T. M. Rice, S. Gopalan and M. Sigrist, unpublished.
\item A. J. Millis and H. Monien, {\em Phys. Rev. Lett.} {\bf 70}, 2810
(1993).
\item M. Arjunwadkar, G. Baskaran, R. Basu and V. N. Muthukumar, {\em Phys.
Rev. Lett.} {\bf 70}, 674 (1993).
\item J. A. Riera, in preparation.
\item It is instructive to see that for $t_{\perp} = 0$,
this kind of interlayer phase separation also occurs when
the total system consists of three or four  $4 \times 4$
clusters.
\item F. F. Assaad and D. Wurtz, {\em Phys. Rev. } {\bf B 44}, 2681
(1991).
\item R. M. Fye, D. J. Scalapino and R. T. Scalettar, {\em Phys. Rev.}
{\bf B46}, 8667 (1992).
\end{enumerate}

\newpage
\centerline {\bf FIGURE CAPTIONS}
\vskip 2truecm

\noindent
{\bf Figure 1}
\noindent
a) Interlayer density fluctuations, b) intralayer PS order parameter,
and c) pairing susceptibility, as a function of $\Jparallel$ for the
coupled 8 sites chain with four holes. The symbols corresponding to
various
values of $\tperp$ are defined in a). In d), the pairing correlations
computed at the peak of the susceptibility as a function of the
distance are shown.
\vskip 1truecm


\noindent
{\bf Figure 2}
\noindent
a) Interlayer density fluctuations, b) intralayer PS order parameter,
and c) pairing susceptibility, as a function of $\Jparallel$ for the
coupled 8 sites chain with 8 holes.
d) Pairing correlations computed
at the peak of the susceptibility as a function of the distance.
The symbols for different $\tperp$ are defined as in Fig. 1.
\vskip 1truecm

\noindent
{\bf Figure 3}
\noindent
a) Interlayer density fluctuations, b) intralayer PS order parameter,
and c) pairing susceptibility, as a function of $\Jparallel$ for the
coupled $\sqrt{8} \times \sqrt{8}$ planes with 8 holes.
d) Pairing correlations computed
at the peak of the susceptibility as a function of the distance for
8 holes, $\tperp$ = 0.0 (circles), 0.6 (squares), 1.0 (diamonds),
and 4 holes, $\tperp$ = 0.0 (filled circles), 0.4 (filled squares).
\vskip 1truecm

\end{document}